\documentclass{article}

\usepackage{amsmath,amsfonts,bm}
\usepackage{amsthm}

\newcommand{\diag}[1]{\operatorname{diag}\left( #1\right)}

\def\eqref#1{equation~\ref{#1}}

\def\1{\bm{1}}

\def\rvf{{\mathbf{f}}}
\def\rvg{{\mathbf{g}}}

\def\rvr{{\mathbf{r}}}
\def\rvs{{\mathbf{s}}}

\def\rmD{{\mathbf{D}}}

\def\rmI{{\mathbf{I}}}

\def\rmR{{\mathbf{R}}}

\def\rmW{{\mathbf{W}}}

\def\mSigma{{\bm{\Sigma}}}

\DeclareMathAlphabet{\mathsfit}{\encodingdefault}{\sfdefault}{m}{sl}
\SetMathAlphabet{\mathsfit}{bold}{\encodingdefault}{\sfdefault}{bx}{n}

\newcommand{\R}{\mathbb{R}}

    \usepackage[preprint]{neurips_2023}

\usepackage[utf8]{inputenc} 
\usepackage[T1]{fontenc}    
\usepackage{hyperref}       
\usepackage{url}            
\usepackage{booktabs}       
\usepackage{amsfonts}       
\usepackage{nicefrac}       
\usepackage{microtype}      

\usepackage[belowskip=-12pt,aboveskip=12pt]{caption}

\usepackage{siunitx}
\usepackage[left]{lineno}
\usepackage{chngcntr}

\usepackage{times}
\usepackage{amsmath}
\usepackage{amssymb}
\usepackage{graphicx}
\usepackage[dvipsnames]{xcolor}
\usepackage{color}
\usepackage{soul}
\hypersetup{
    colorlinks=true,
    linkcolor=MidnightBlue,
    filecolor=magenta,      
    urlcolor=ForestGreen,
    citecolor=Violet,
    }

\setcitestyle{apalike,round}
\title{Adaptive Coding Efficiency in Recurrent \\ Cortical Circuits via Gain Control}

\author{%
  Lyndon R. Duong\textsuperscript{1}, Colin Bredenberg\textsuperscript{1}, David J. Heeger\textsuperscript{1,2}, Eero P. Simoncelli\textsuperscript{1,3} \\
  \textsuperscript{1}Center for Neural Science, New York University, New York, NY\\
  \textsuperscript{2}Department of Psychology, New York University, New York, NY\\
  \textsuperscript{3}Center for Computational Neuroscience, Flatiron Institute, New York, NY\\
  \texttt{\small \{lyndon.duong, cjb617, david.heeger, eero.simoncelli\}@nyu.edu}
  }

\begin{document}

\maketitle

\begin{abstract}
Sensory systems across all modalities and species exhibit adaptation to continuously changing input statistics.
Individual neurons have been shown to modulate their response gains so as to maximize information transmission in different stimulus contexts.
Experimental measurements have revealed additional, nuanced sensory adaptation effects including changes in response maxima and minima, tuning curve repulsion from the adapter stimulus, and stimulus-driven response decorrelation.
Existing explanations of these phenomena rely on changes in inter-neuronal synaptic efficacy, which, while more flexible, are unlikely to operate as rapidly or reversibly as single neuron gain modulations.
Using published V1 population adaptation data, we show that propagation of single neuron gain changes in a recurrent network is sufficient to capture the entire set of observed adaptation effects.
We propose a novel adaptive efficient coding objective with which single neuron gains are modulated, maximizing the fidelity of the stimulus representation while minimizing overall activity in the network.
From this objective, we analytically derive a set of gains that optimize the trade-off between preserving information about the stimulus and conserving metabolic resources.
Our model generalizes well-established concepts of single neuron adaptive gain control to recurrent populations, and parsimoniously explains experimental adaptation data.
\end{abstract}



\section{Introduction}

Some of the earliest neurophysiological recordings showed that repeated or prolonged stimulus presentation leads to a relative decrease in neural responses \citep{adrian_impulses_1926}. 
Indeed, neurons across different species, brain areas, and sensory modalities adjust their gains (i.e. input-output sensitivity) in response to recent stimulus history \citep[][for reviews]{kohn_visual_2007, weber_coding_2019}.
Gain control provides a mechanism for single neurons to rapidly and reversibly adapt to different stimulus contexts \citep{abbott_synaptic_1997, brenner_adaptive_2000, fairhall_efficiency_2001,  mlynarski_efficient_2021, muller1999rapid} while preserving synaptic weights that serve to represent features that remain consistent across contexts \citep{ganguli_efficient_2014}.
From a normative standpoint, this allows a  single neuron to adjust the dynamic range of its responses to accommodate changes in input statistics  \citep{laughlin_simple_1981, fairhall_efficiency_2001} -- a core tenet of theories of efficient sensory coding \citep{attneave_informational_1954, barlow_possible_1961}.

Experimental measurements, however, reveal that adaptation induces additional complex changes in neural responses, including tuning-dependent reductions in both response maxima and minima \citep{movshon_pattern-selective_1979}, tuning curve repulsion \citep{shen_frequency-specific_2015, hershenhoren_intracellular_2014, yaron_sensitivity_2012}, and stimulus-driven decorrelation \citep{benucci_adaptation_2013, wanner_whitening_2020,gutnisky_adaptive_2008,muller1999rapid}.
Although coding efficiency and gain-mediated adaptation is well studied in single neurons, it appears as though these nuanced empirical observations require a more complex adaptation mechanism, involving \textit{joint} coordination among neurons in the population.
Indeed, to explain these phenomena, previous studies have relied on adaptive changes in feedforward or recurrent synaptic efficacy \citep[i.e. by changing the entire network's set of synaptic weights;][]{wainwright_natural_2001, westrick_pattern_2016, rast_adaptation_2021,mlynarski_efficient_2021}.
However, this requires synaptic weights to continuously remap under different statistical contexts, which may change significantly and transiently at short time scales.

Here, we hypothesize that adaptation effects reported in neural population recording data can be explained by combining normative theory with a mechanistic recurrent population model that includes single neuron gain modulation. 
The primary contributions of our study are as follows:
\begin{enumerate}
    \item We introduce an analytically tractable recurrent neural network (RNN) architecture for adaptive gain control, in which single neurons adjust their gains in response to novel stimulus statistics.
    The model respects experimental evidence that cortical anatomy is dominated by recurrence \citep{douglas_recurrent_2007}, allowing the effects of single neuron gain changes to propagate through lateral connections.
    \item  We propose a novel \textit{adaptive efficient coding} objective for adjustment of the single neuron gains, which optimizes coding fidelity of the stimulus ensemble, subject to metabolic and homeostatic constraints.
    \item  Through numerical simulations, we compare model predictions to experimental measurements of cat V1 neurons responding to a sequence of gratings drawn from an ensemble with either uniform or biased orientation probability \citep{benucci_adaptation_2013}.
    We show that adaptive adjustment of neural gains, with no changes in synaptic strengths, parsimoniously captures the full set of adaptation phenomena observed in the data.
\end{enumerate}

\begin{figure}[t]
\centering
\includegraphics[width=\textwidth]{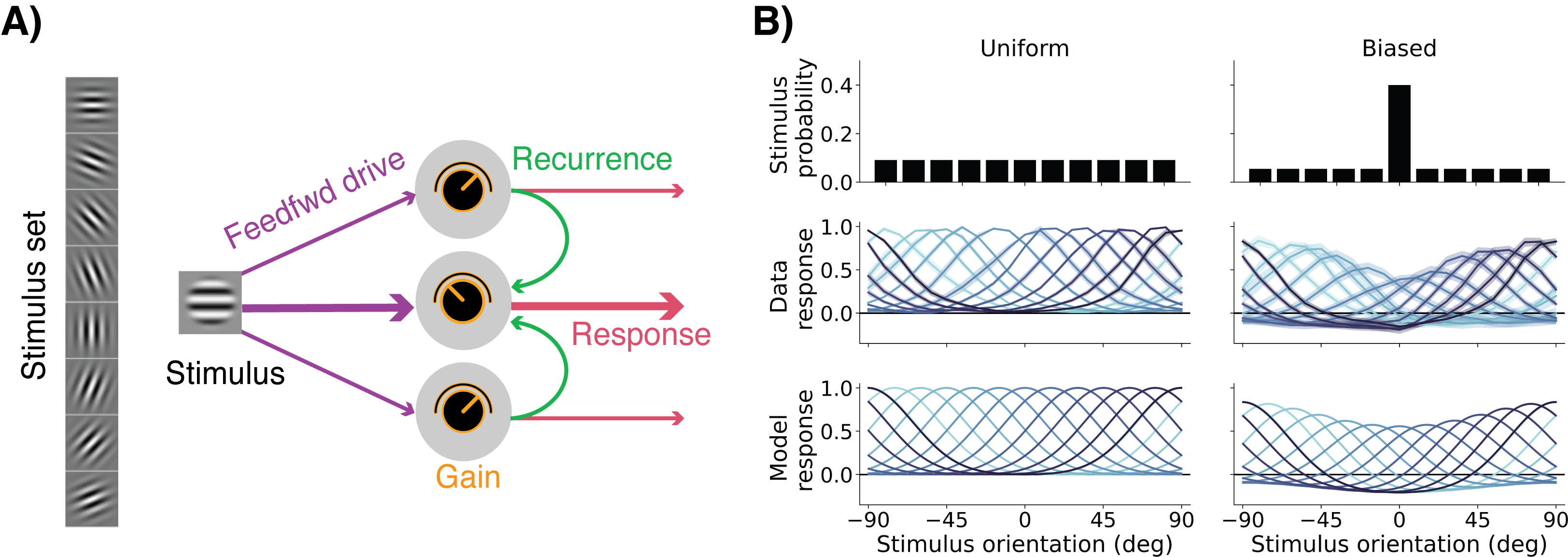}
 \caption{\small Recurrent adaptation model.
{\bf A)} A population of recurrently-connected orientation-tuned cells receives external feedforward drive (purple arrows) from a presented oriented grating stimulus, randomly sampled from a set of possible orientations. The width of the arrow denotes the strength of the drive, and indicates that the center neuron is tuned towards the horizontal-oriented stimuli. The feedforward drive of each neuron is multiplicatively modulated by its a scalar gain (orange dials). Lateral recurrent input between neurons is denoted by green arrows. Recurrent connectivity is all-to-all, with synaptic strengths determined by the distance between neurons' preferred feedforward orientation. Output responses (red) of each neuron are a function of both feedforward drive and recurrent drive.
{\bf B)} Response tuning curves for orientation-tuned units to stimuli presented with uniform probability (left column), or biased probability (right column).  Middle row shows recordings of neurons in visual area V1 of cats, aggregated over 11 sessions.  Bottom row shows model responses.  
Shaded regions are standard error of the mean (SEM).
 }
 \label{fig:model}
\end{figure}

\section{Related Work}

\paragraph{Models of statistical adaptation in neural populations.}
While evidence for adaptive efficient coding via gain modulation in single neurons is relatively well understood \citep{fairhall_efficiency_2001,nagel_temporal_2006,mlynarski_efficient_2021}, the question of whether neural \textit{population} adaptation can be explained by efficient coding and gain modulation remains underexplored.
Normative models of population adaptation have generally relied on synaptic plasticity (i.e. between-neuron synaptic weight adjustments) as the mechanism mediating adaptation \citep{mlynarski_efficient_2021, rast_adaptation_2021, pehlevan2015normative, lipshutz2022interneurons, wainwright_natural_2001,westrick_pattern_2016}.
For example, \citet{westrick_pattern_2016} argue that empirical observations of V1 neural populations \citep{benucci_adaptation_2013} can be explained by adapting normalization weights (parameterized by all-to-all synaptic connections) to different stimulus statistical contexts.
The major downside of this approach is that changes in synaptic weights require $\mathcal{O}(N^2)$ adaptation parameters, for a population of size $N$.
Here, we examine the effects of classical single-neuron adaptive gain modulation on responses of a recurrently-connected population, and demonstrate that these are sufficient to explain adaptation phenomena, while requiring only $\mathcal{O}(N)$ adaptation parameters.
Holding the synaptic weights fixed prevents overfitting, and allows the network to remain stable across input contexts.
Network stability is also relevant for contemporary machine learning applications that rely on adaptive adjustments to changing input statistics  \citep[e.g.][]{hu2021lora,balle_nonlinear_2020,mohan_adaptive_2021}.

The adaptation model most similar to ours,  developed by \citet{gutierrez_population_2019}, proposes an adaptive recurrent spiking neural network whose dynamics are derived from an efficient coding objective.
Our model is complementary to this, but is simpler and more tractable, providing an analytic solution for population steady-state responses that facilitates comparisons to experimental data. 
Finally, recent work (published while this manuscript was being written) uses gain control as a normative population adaptation mechanism, but with the central goal of statistically whitening neural responses, while ignoring the means of responses \citep[i.e. redundancy reduction via decorrelation and variance equalization; ][]{duong2023statistical}.
Here, we demonstrate that our model captures adaptive effects involving mean responses as well as population response redundancy reduction, but that its steady-state responses are not whitened. 
We show that these deviations from whitening are similar to those seen in the neural recordings analyzed here.

\paragraph{Recurrent circuitry in sensory cortex.} It is well known that recurrent excitation dominates cortical circuits \citep{douglas_recurrent_2007}.
In early sensory areas, a series of optogenetic inactivation experiments showed that recurrent excitation in cortex serves to progressively amplify thalamic inputs \citep{reinhold_distinct_2015, lien2013tuned}.
In the context of sensory adaptation, \citet{king_adaptive_2016} performed silencing experiments in mice to show that the majority of adaptation effects seen in V1 arise from \textit{local} activity-dependent processes,  rather than being inherited from depressed thalamic responses upstream.
Similarly, in monkey V1 neurophysiological recordings, \citet{westerberg2019v1} used current source density analyses to show that stimulus-driven adaptation is primarily due to recurrent intracortical effects rather than feedforward effects.
We leverage these functional observations, along with anatomical measurements of intracortical synaptic connectivity \citep{ko2011functional,lee2016anatomy,rossi2020spatial} to inform the recurrent architecture used in our study.

\section{An Analytically Tractable RNN with Gain Modulation}

\subsection{Notation}
We denote matrices with capital boldface letters (e.g. $\rmW$), vectors as lowercase boldface letters (e.g. $\rvr$), and scalar quantities as non-boldface letters (e.g. $N, \alpha$). 
The $\operatorname{diag(\cdot)}$ operator forms a diagonal matrix by embedding the elements of a $K$-dimensional vector onto the main diagonal of a $K \times K$ matrix whose off-diagonal elements are zero.
$\circ$ is the Hadamard (i.e. element-wise) product.
$\mathbb{S}_+^N$ is the space of $N \times N$ symmetric positive definite matrices.

\subsection{Adaptive gain modulation in a population without recurrence}

We first consider the steady-state response  of $N$ neurons, $\rvr_{\text{f}} \in \mathbb{R}^N$, receiving sensory stimulus inputs $\rvs \in \R^M$, with feedforward drive, ${\bf f(s)} = [f_1(\rvs), f_2(\rvs), \dots, f_N(\rvs)]^\top$, which are each multiplicatively scaled by gains, $\rvg = [g_1, g_2, \dots, g_N]^\top$: 
\begin{equation}
    {\bf r_{\text{f}}}(\rvs, \rvg) = \rvg \circ \rvf (\rvs). \label{eq:ffwd_only}
\end{equation}
The gain vector $\rvg$ has the effect of adjusting the amplitudes of responses ${\bf f(s)}$, and therefore the dynamic range of each neuron.
As we demonstrate in Section~\ref{sec:results}, these simple multiplicative gain scalings are incapable of shifting the peaks of tuning curves, as seen in physiological data~\citep{movshon_pattern-selective_1979, muller1999rapid,saul1989adaptation}.
Previous approaches modeling neural population adaptation in cortex modify the structure of ${\bf f(s)}$ in response to changes in input statistics \citep[e.g.][]{wainwright_natural_2001, westrick_pattern_2016}. Here, we propose a fundamentally different approach, requiring \textit{no} changes in synaptic weighting between neurons.

\subsection{Gain modulation in a recurrent neural population}
We show that by incorporating single neuron gain modulation into a recurrent network, adaptive effects in each neuron propagate laterally to affect other cells in the population.
Consider a model of $N$ \textit{recurrently} connected neurons with fixed feedforward and recurrent weights (Fig. \ref{fig:model}A), presumed to have been learned over timescales much longer than the adaptive timescales examined in this study.
We assume that the population of neural responses $\rvr \in \mathbb{R}^N$, driven by input stimuli $\rvs \in \mathbb{R}^M$ presented with probability $p(\rvs)$, are governed by linear dynamics:
\begin{equation}
     \frac{d\rvr(\rvs, \rvg)}{dt}= -\rvr + \rvg\circ\rvf(\rvs) + \rmW  \rvr, \label{eq:drdt_linear}
\end{equation}
where $\rmW \in \mathbb{R}^{N\times N}$ is a matrix of recurrent synaptic connection weights; and neuronal gains, $\rvg \in \mathbb{R}^N$, are adaptively optimized to a given $p(\rvs)$.
Both the feedforward functions $f_i(\rvs)$ and recurrent weights $\rmW$ are assumed to be fixed despite varying stimulus contexts (i.e. \textit{non-adaptive}).
For notational convenience, we omit explicit time-dependence of the responses and stimuli (i.e. $\rvr(\rvs, \rvg, t), \rvs(t)$).

Empirical studies typically consider neural activity at steady-state before and after adapting to changes in stimulus statistics \citep[][]{clifford_visual_2007}. 
We therefore analyze the responses of our network at steady-state, ${\bf r_\ast(s, g)}$, to facilitate comparison with data.
The network dynamics of Equation~\ref{eq:drdt_linear} are linear in $\rvr$, and computing its steady-state is analytically tractable.
Setting Eq.~\ref{eq:drdt_linear} to zero and isolating $\rvr$ (with the mild assumptions on invertibility; see Appendix~\ref{appendix:W}), yields the steady-state solution, 
\begin{equation}
    {\rvr_{\ast}}(\rvs, \rvg) = \left[\rmI- \rmW \right]^{-1} \left(\rvg \circ \rvf(\rvs)\right). \label{eq:ss}
\end{equation}
We can interpret these equilibrium responses as a modification of the gain-modulated feedforward drive, $\rvg \circ \rvf(\rvs)$, which is propagated to other cells in the network via recurrent interactions,  $\left[\rmI - \rmW \right]^{-1}$.
When $\rmW$ is the zeros matrix (i.e. no recurrence), Equation~\ref{eq:ss} reduces to Equation~\ref{eq:ffwd_only}, and adjusting neuronal gains simply rescales the feedforward responses without affecting the shape of response curves. The presence of the recurrent weight matrix $\rmW$ allows changes in neuronal gains to alter the effective tuning of other neurons in the network \textit{without} changes to any synaptic weights.

\subsection{Structure of recurrent connectivity matrix $\rmW$}

Importantly, in our recurrent network, there are no explicit excitatory and inhibitory neurons -- the recurrent activity term (last term in Eq.~\ref{eq:drdt_linear}) represents the \textit{net} lateral input to a neuron (i.e. the combination of both excitatory and inhibitory inputs).
In addition, model simulations in this study use a $\rmW$ that is translation invariant (i.e. convolutional) in preferred orientation space, with strong net recurrent excitation near the preferred orientation of the cell, and relatively weak net excitation far away.
This structure is motivated by functional and anatomical measurements in V1, indicating that orientation-tuned cells receive excitatory and inhibitory presynaptic inputs from cells tuned to every orientation, with disproportionate excitatory bias from similarly-tuned neurons \citep{lee2016anatomy,rossi2020spatial,rubin_stabilized_2015}.
We elaborate on specific choices of $\rmW$ in Appendix~\ref{appendix:W}.

\section{A Novel Objective for Adaptive Efficient Coding via Gain Modulation}

Theories of efficient coding postulate that sensory neurons optimally encode the statistics of the natural environment \citep{barlow_possible_1961, laughlin_simple_1981}, subject to constraints on finite metabolic resources \citep[e.g. energy expenditure from firing spikes;][]{ganguli_efficient_2014,olshausen_emergence_1996}.
However, sensory input statistics vary with context, and the means by which a neural population might confer an \textit{adaptive and dynamic} efficient code remains an open question \citep{gutierrez_population_2019,mlynarski_efficient_2021,barlow_adaptation_1989,duong2023statistical}.
How should our network (Equation~\ref{eq:ss}) adaptively modulate its gains, $\rvg$, according to the statistics of a novel stimulus ensemble?
We assume an initial stimulus ensemble, with probability density $p_0(\rvs)$ (Fig. \ref{fig:model}B), with a corresponding set of optimal gains, $\rvg_0$, toward which adaptive gains are homeostatically driven; and an optimal linear decoder, $\rmD \in \mathbb{R}^{N\times M}$.
$\rmD$ is fixed and set to the pseudoinverse of $\rvr_\ast(\rvg, \rvs)$ under the initial stimulus ensemble (see Appendix~\ref{appendix:isolating_g}).

Given a novel stimulus ensemble with probability density $p(\rvs)$, we propose an adaptive efficient coding objective that neurons minimize by adjusting their gains,
\begin{equation}
\mathcal{L}(\rvg, p(\rvs)) =  \mathbb{E}_{\rvs \sim p(\rvs)} \left \{\Vert \rvs  - \rmD^\top \rvr_\ast (\rvs, \rvg) \Vert_2^2 + \alpha \Vert \rvr_\ast (\rvs, \rvg) \parallel_2^2 \right\} + \gamma  \parallel \rvg - \rvg_0\parallel^2_2 \label{eq:objective},  
\end{equation}  
where $\alpha$ and $\gamma$ are scalar hyperparameters.
Intuitively, as the stimulus ensemble changes $ p_0(\rvs) \rightarrow p(\rvs)$, the gains $\rvg$ are adaptively adjusted to maximize the fidelity of the representation (first term), while minimizing overall activity in the network (second term), and minimally deviating from the  initial gain state (third term).
The gain homeostasis term serves to prevent catastrophic forgetting in the network under different stimulus contexts \citep{kirkpatrick2017overcoming}: minimizing the gains' deviation from their optimal state under $p_0(\rvs)$ allows the system to stably maintain reasonable performance on previously presented data and prevents the system from radically reorganizing itself on a fast time scale. 
In Appendix~\ref{appendix:ablations}, we show that adapting to $p(\rvs)$ with gain homeostasis allows the network to maintain improved stimulus representation error under the $p_0(\rvs)$ ensemble relative to a network optimized without gain homeostasis.
We also perform ablations to show that the three terms in the objective are \textit{jointly} necessary to produce the adaptation effects observed in data.

\subsection{Objective optimization}
The objective given in Equation~\ref{eq:objective} is bi-convex in $\rvg$ and $\rmD$, and we can \textit{analytically} solve for either variable independently or in alternation (i.e., coordinate descent via alternating least squares).
See Appendix~\ref{appendix:isolating_g} for the complete derivation.
We initialize the network under the uniform stimulus density $p_0(\rvs)$ to obtain a homeostatic gain target, $\rvg_0$, and a fixed decoder, $\rmD$.

\section{V1 Neural Population Adaptation Data Reanalysis}
\label{sec:cat_data}

In the following section, we compare our simulated adaptation model responses to reanalyzed neural population recordings from cat primary visual cortex \citep[data obtained with permission from][]{benucci_adaptation_2013}.
Here, we provide an overview of our data analysis procedure which we also apply to our simulated model responses.
Some of our analysis plots are new and are not in the original study\footnote{Additionally, our plots are derived from steady-state fitted response curves, whereas the original publication used temporal information.}.
For details on the recordings and preprocessing, we refer the reader to the original paper.

In the experiment, oriented stimuli were briefly presented randomly in rapid succession, with presentation probability determined by one of two contextual distributions: a 
uniform distribution $p_0(\rvs)$, or a {\em biased} distribution, in which one orientation was presented significantly more frequently than the others, $p(\rvs)$ (Figure~\ref{fig:model}B, top row).
Figure~\ref{fig:model}B (middle row) shows responses for $N=13$ units, aggregated over 11 recording sessions.
For $N$ units and $K$ distinct stimuli, the authors fit orientation tuning curves to neural responses to produce matrices of orientation tuning curves, $\rmR \in \mathbb{R}^{N\times K}$ for each of the uniform and biased stimulus ensembles.

We normalize each unit's response curves under both contexts according to its minimum and maximum response during the $p_0(\rvs)$ context, such that all responses lie in the interval [0, 1] for $p_0(\rvs)$. That is, zero is the minimum stimulus-evoked response under the uniform ensemble, and one is the maximum.
For responses to the biased ensemble, $p(\rvs)$, a minimum response less than 0 indicates that the evoked response after adaptation has decreased relative to the uniform ensemble; similarly, a  maximum response less than 1 indicates the response maximum after adaptation has decreased relative to that of the uniform ensemble (Figure~\ref{fig:model}B).

We compute response means, $\boldsymbol{\mu}  \in \mathbb{R}^N$, and signal (as opposed to noise) covariance matrices, $\mSigma \in \mathbb{S}_{+}^N$, 
\begin{equation}
        \boldsymbol{\mu} = \mathbb{E}[\rmR],
    \qquad \mSigma = \mathbb{E}[\rmR \rmR^\top] - \boldsymbol{\mu}\boldsymbol{\mu}^\top,
\end{equation}
where the expectation is over $p_0(\rvs)$ or $p(\rvs)$.
To facilitate comparisons between response covariances under the uniform and biased stimulus ensembles, we scale response covariance matrices by the variances of the neurons under the uniform stimulus probability condition, ${\bm \sigma}_{0}^2 \in \mathbb{R}_+^N$,
\begin{equation}\label{eq:scaled_sigma}
    \hat\mSigma = \diag{{\bm \sigma}_{0}}^{-1} \mSigma \diag{{\bm \sigma}_{0}}^{-1}.
\end{equation}


\section{Numerical Simulations and Comparisons to Neural Data}
\label{sec:results}

We compare numerical simulations of our normative adaptation model with reanalyzed cat V1 population recording data \citep{benucci_adaptation_2013}.

\subsection{Model and simulation parameters}

For all simulation results and figures in this study, we consider a network comprised of $N=255$ recurrently connected neurons, with $K=M=511$ orientation stimuli as inputs.
The neuronal gains, $\rvg$, adapt to changes in stimulus ensemble statistics ($p_0(\rvs) \rightarrow p(\rvs)$), while the feedforward synaptic weights, $\rvf(\rvs)$, and recurrent synaptic weights, $\rmW$, remain fixed.
We set the homeostatic target gains, $\rvg_0$,  to the optimal values of $\rvg$ under the uniform probability stimulus ensemble, $p_0(\rvs)$.
Feedforward orientation-tuning functions, $\rvf(\rvs)$, are evenly distributed in the stimulus domain, and are broadly-tuned Gaussians with full-width half-max (FWHM) of 30\textsuperscript{$\circ$} \citep{benucci_adaptation_2013}.
The recurrent weight matrix, $\rmW$, is a Gaussian with 10\textsuperscript{$\circ$} FWHM, summed with a weaker, broad, untuned excitatory component (see Appendix \ref{appendix:W}).

To determine appropriate values of $\alpha$ and $\gamma$ in the objective (Equation~\ref{eq:objective}), we performed a grid-search hyperparameter sweep, minimizing the deviation between model and experimentally-measured tuning curves for the biased stimulus ensemble.
The figures here all use model responses from a simulation using $\alpha=$ 1E-3, $\gamma=$ 1E-2.
We find that qualitative effects are insensitive to small changes in these parameters.
The key finding from this parameter sweep is that the gain homeostasis penalty weight must be sufficiently greater than the activity penalty weight (i.e. $\gamma > \alpha$).
After initializing the network gains to the statistics of $p_0(\rvs)$, we adapt the gains to $p(\rvs)$ by optimizing Equation~\ref{eq:objective}, then compare our model predicted responses to cat V1 population recordings Figure~\ref{fig:model}B. 

\begin{figure}[t]
\centering
\includegraphics[width=\textwidth]{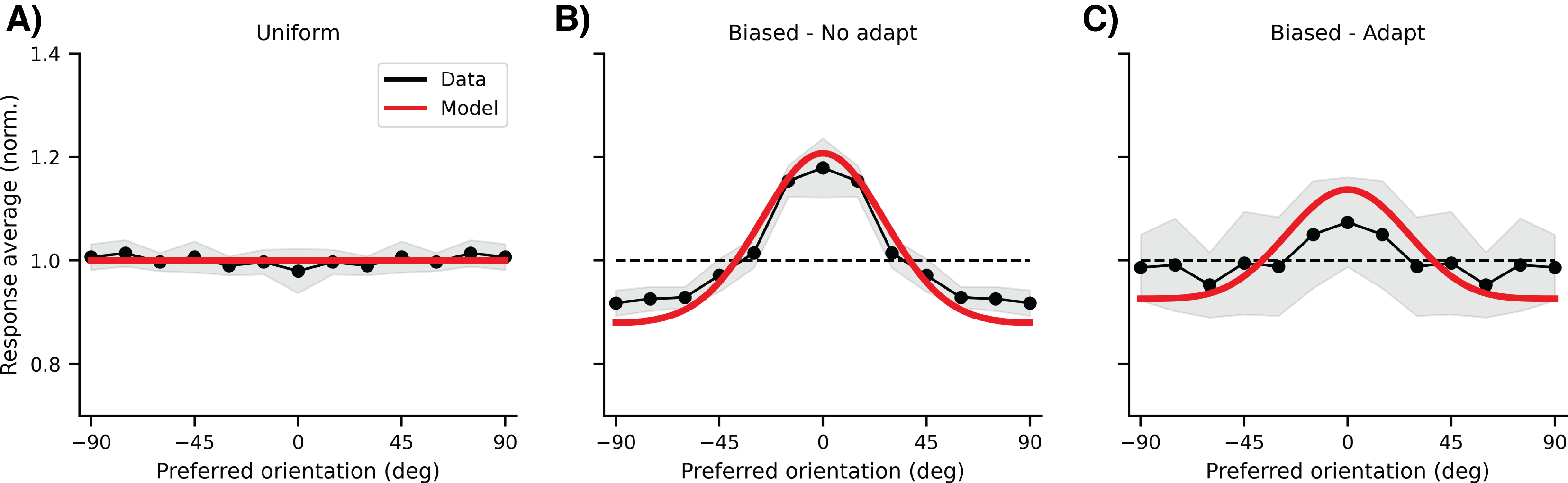}
 \caption{\small 
 Adaptive response equalization. Each dot is the average response of a neuron. 
 {\bf A) } Response averages under the uniform stimulus ensemble condition.
 {\bf B) } \textit{Without} adaptation, response averages under the biased stimulus ensemble show substantial deviation from equalization (which corresponds to the dashed black line).
 {\bf C) } After adaptation, response averages to the biased ensemble are nearly equalized.
 Shaded regions are SEM.
 }
 \label{fig:fr_mean}
\end{figure}

\subsection{Adaptive gain modulation predicts response equalization}

Population response equalization is an adaptive mechanism first proposed in the psychophysics literature \citep{anstis1998motion}.
The authors argued that adaptation should serve as a ``graphic equalizer'' in response to alterations in environmental statistics. 
Others have have described equalization as a mechanism that centers a population response by subtracting the responses to the prevailing stimulus ensemble \citep{clifford2000functional}, to rescale responses such that the average of a measured signal remains constant \citep{ullman1982adaptation}.
Figure~\ref{fig:fr_mean} shows how our model recapitulates mean firing rates across all stimuli under the uniform and biased ensembles without adaptation, along with adaptive population response equalization under the biased ensemble.
Figure~\ref{fig:fr_mean}A shows how the average response of each pre-adapted neuron under the uniform ensemble is equal.
By contrast, Figure~\ref{fig:fr_mean}B demonstrates that our model predicts how the pre-adaptation tuning curves under the biased stimulus ensemble would produce a substantial deviation from equalization.
Finally, adaptively optimizing neuron gains via Equation~\ref{eq:objective} predicts the compensatory response equalization under the biased stimulus ensemble observed in data (panel C).

\begin{figure}[htb]
\centering
\includegraphics[width=\textwidth]{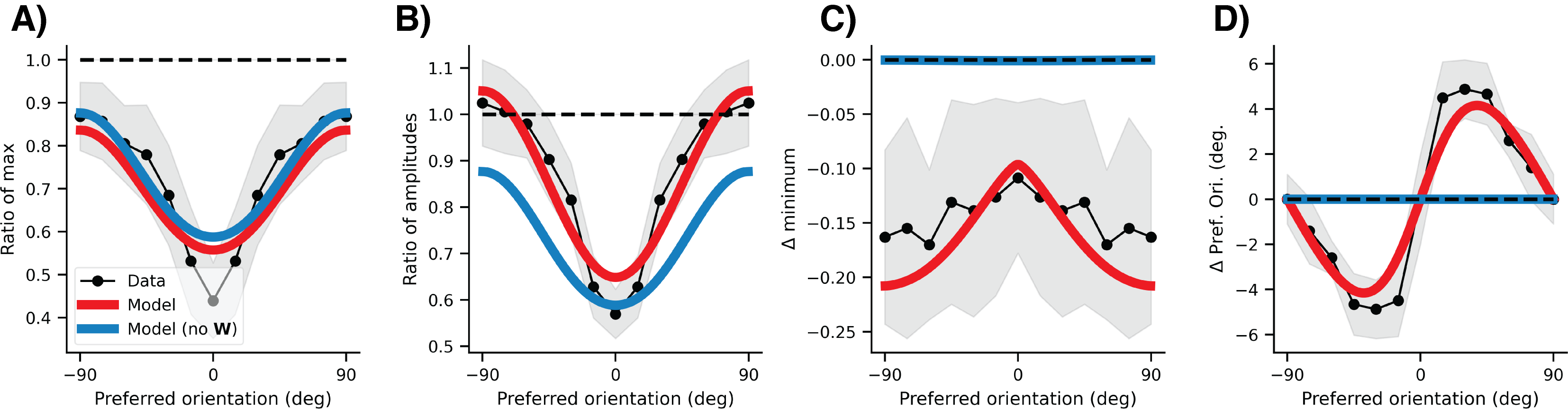}
 \caption{\small
 Recurrent network model with adaptive gain modulation (red) captures the full set of post-adaptation first-order response changes observed in data (black points), while a network without recurrence (blue) does not.
 {\bf A)} Ratios of after/before adaptation response maxima. 
 {\bf B)} Ratios of response amplitudes ($\Vert \max - \min \Vert$ response) after/before adaptation.
 {\bf C)} Changes in average minimum evoked response.
 {\bf D)} Shifts in tuning away from the adapter.
Shaded regions are SEM.
Dashed lines indicate predictions for a non-adaptive model.
 }
 \label{fig:tc_changes}
\end{figure}

\subsection{Adaptive gain modulation predicts nuanced changes in first-order statistics of responses}

Figure~\ref{fig:tc_changes} summarizes adaptive changes in neural responses by comparing tuning curve responses under the biased stimulus ensemble compared to responses under the uniform stimulus ensemble (i.e. right vs. left columns of Fig.~\ref{fig:model}B).
Our gain-modulating efficient coding model can capture this entire array of observed adaptation effects.

\paragraph{Changes in response maxima, amplitudes, and minima.}
Stimulus-dependent response reductions are a ubiquitous finding in adaptation experiments \citep{weber_coding_2019}.
Figure~\ref{fig:tc_changes}A,B show changes in response maxima, and response amplitudes (peak-to-trough height) following adaptation to the biased stimulus ensemble.
Ratios less than 1 indicate a reduction in maxima or amplitudes following adaptation.
Under the biased stimulus ensemble, the model optimizes its gains according to the objective (Eq.~\ref{eq:objective}) to preferentially reduce activity near the over-presented adapter stimulus.
By optimizing gains to the adaptive efficient coding objective, our linear model undershoots the magnitude of change around the adapter (Fig.~\ref{fig:tc_changes}A,B red curve near 0\textsuperscript{$\circ$}), but captures the overall effect of adaptive amplitude and maxima reduction.

Figure~\ref{fig:tc_changes}C shows that adaptation induces a tuning-dependent, global reduction in minimum stimulus-evoked response across the population.
These minima typically occur at the anti-preferred orientation for each neuron (Fig.~\ref{fig:model}B).
Previous work has attributed this to an untuned reduction in thalamic inputs, or a drop in base firing \citep{benucci_adaptation_2013,westrick_pattern_2016}.
Our model proposes a different mechanism: gain reductions in neurons tuned for the adapter propagate laterally through the network, and result in commensurate reductions in the broad/untuned recurrent excitation to other neurons in the population.
This ultimately leads to a reduction in minimum evoked response across the entire population (Fig.~\ref{fig:tc_changes}C); importantly, the model also captures the qualitative shape of the change.
Our mechanistic prediction that this effect arises due to recurrent contributions is in concordance with the broad literature on recurrent cortical circuitry, its role in amplification \citep{reinhold_distinct_2015}, and in sensory adaptation \citep{hershenhoren_intracellular_2014,king_adaptive_2016}.

\paragraph{Shifts in tuning preference.}
Tuning curve shifts following adaptation have been reported across many visual and auditory adaptation studies \citep[][for reviews]{clifford_visual_2007,whitmire_rapid_2016}.
Figure~\ref{fig:tc_changes}D quantifies changes in neuron preferred orientation (i.e. the orientation at which response maximum occurs) after adapting to the biased stimulus ensemble.
The sinusoidal shape of the curve indicates that adapted tuning curves are \textit{repelled} from the over-presented adapter stimulus.
This rearrangement of tuning curve density is consistent with efficient coding studies that argue that a sensory neural population should optimally allocate its finite resources toward encoding    information about the current stimulus ensemble \citep{gutnisky_adaptive_2008,ganguli_efficient_2014}.
Here, we show that these effects can mechanistically be explained by optimizing neuronal gains to maintain a high fidelity representation of the stimulus under the biased ensemble.

\paragraph{Objective and network ablations.}
In Appendix~\ref{appendix:ablations}, we assess the importance of each term of the adaptation objective (Eq.~\ref{eq:objective}) by ablating them from the objective and re-simulating the network adapting to the biased stimulus ensemble.
We show that each of the three terms is jointly necessary to capture the adaptation effects shown here.
In terms of network architectural ablations, the blue curves in Figure~\ref{fig:tc_changes} demonstrate how removing recurrence (i.e. $\rmW= 0$; Equation~\ref{eq:ffwd_only}) impacts adaptive changes in neural responses.
While this single stage feedforward model can reproduce reductions in response maxima Fig.~\ref{fig:tc_changes}A, it is incapable of producing the appropriate change in response amplitudes (Fig.~\ref{fig:tc_changes}B), and completely fails at producing adaptive reductions in minimum response, or shifts in tuning preference (Fig.~\ref{fig:tc_changes}C,D). 
Intuitively, this is because the gains in this reduced model serve to set the amplitude of the output, and cannot alter the qualitative shape of the tuning curve without propagating through the recurrent circuitry.
The structure of $\rmW$ used in our model is informed by functional and anatomical studies in cortical circuits \citep{lee2016anatomy,rossi2020spatial}, comprising strong net excitation from similarly-tuned neurons and untuned weak net excitation from dissimilarly-tuned neurons.
In Appendix~\ref{appendix:W}, we study the impact of $\rmW$'s structure on model adapted responses.
The structure of $\rmW$ can be quite flexible while still producing the effects shown here, so long as recurrent input includes weak net excitation from dissimilarly-tuned neurons.

\subsection{Adaptive gain modulation predicts homeostasis in second-order statistics of responses}

\begin{figure}[t]
  \centering
\includegraphics[width=\textwidth]{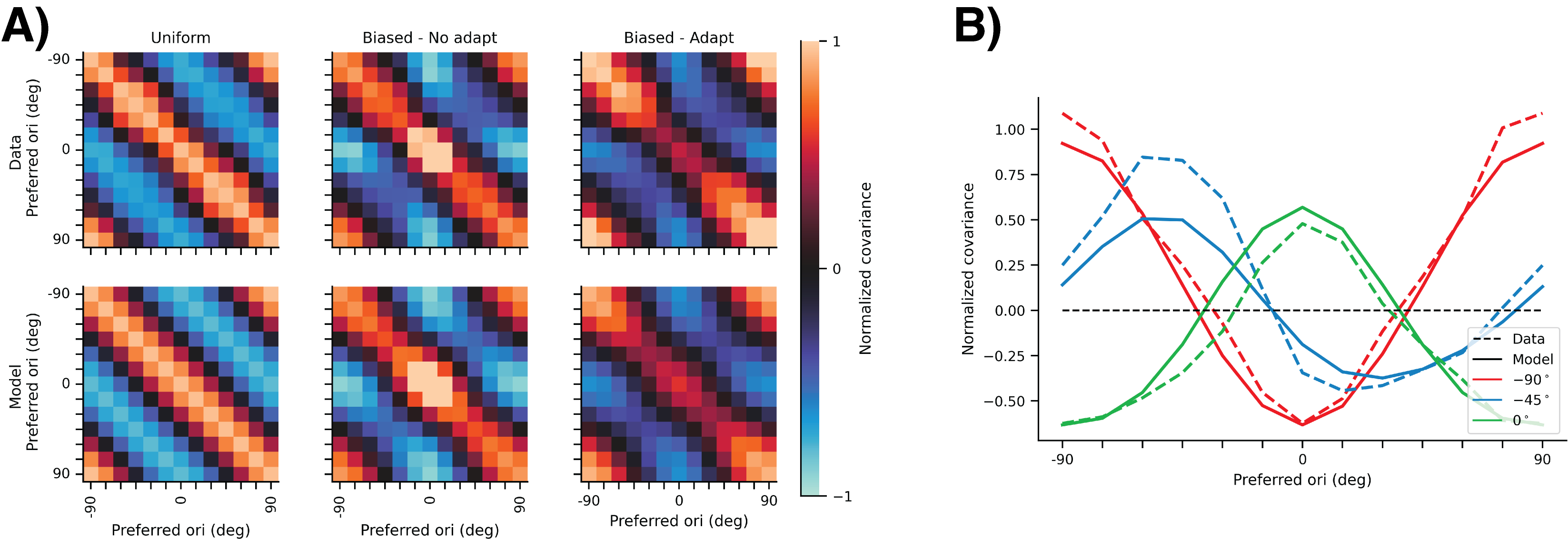}
 \caption{\small
 Population response redundancy reduction and signal covariance homeostasis.
 {\bf A)} Scaled response covariance matrices, $\hat\mSigma$ (Eq.~\ref{eq:scaled_sigma}), for 
 V1 data (top row) and model simulations (bottom row), for unadapted tuning curves and uniform stimulus ensemble (left column), unadapted tuning curves and biased stimulus ensemble (middle column), and adapted tuning curves and biased stimulus ensemble (right column).
 {\bf B)} Three example horizontal slices of the data (dashed) and model (solid) $\hat\mSigma$ from {\bf A}, at 0, -45, and -90 degrees orientation (colors).
 }
 \label{fig:cov}
\end{figure}

The principle of redundancy reduction is core to the efficient coding hypothesis \citep{barlow_possible_1961}, and evidence supporting \textit{adaptive} redundancy reduction has been reported across multiple brain regions and modalities \citep{Atick1992WhatDT,wanner_whitening_2020,muller1999rapid}. 
In the task modeled in our study, over-presenting the adapter stimulus can be viewed as increasing redundancy in the stimulus ensemble (Figure~\ref{fig:model}B, top).
This manifests as a ``hot spot'' in the center of $\hat\Sigma$ if the neural responses were to remain unadapted to $p(\rvs)$ (Fig.~\ref{fig:cov}A, middle column).
However, when the model adapts its gains according to the objective (Eq.~\ref{eq:objective}), the covariance near the adapter stimulus is reduced, and the predicted signal covariance is well matched to data  (Fig.~\ref{fig:cov}A, right column, Fig.~\ref{fig:cov}B).

A signal covariance matrix devoid of redundancy would be one that is statistically white (i.e. the identity matrix).
However, under both the uniform and biased stimulus ensemble conditions (Figure~\ref{fig:cov}A, top left and right), we note that the experimentally observed signal covariance matrix \textit{is not} statistically white\footnote{In their study, \citet[][Fig. 3]{benucci_adaptation_2013} replaced negative entries of $\hat\Sigma$ with zeros.}.
Thus, previous normative approaches to population adaptation that explicitly whiten neural responses may not be suitable models for this data \citep[e.g.][]{pehlevan2015normative, mlynarski_efficient_2021}.
By contrast, our adaptation objective, which emphasizes stimulus signal fidelity subject to metabolic and homeostatic constraints predicts an adapted signal covariance matrix whose deviations from the identity matrix are similar to those observed in data.
Notably, this effect naturally emerges from our model \textit{without} additional parameter-tuning.

\section{Discussion}
\label{sec:discussion}

\paragraph{Study limitations.}
The network considered here is a rate model whose tractable linear dynamics allow us to examine adaptation responses at steady-state.
Response dynamics during adaptation are rich \citep{patterson_distinct_2013,quiroga_adaptation_2016,dragoi2000adaptation}, and are relatively understudied. 
Developing our model and objective into a biologically plausible online network with explicit excitatory and inhibitory neurons, while adapting gains according to only local signals \citep{duong2023statistical,gutierrez_population_2019} is an interesting direction worth pursuing.
Furthermore, because we model trial-averaged experimental data in this study, our model does not account for stochasticity in neural responses. 
Thus, our model cannot explain adaptive changes in trial-to-trial variability \citep{gutnisky_adaptive_2008}.
Finally, there exist adaptive changes to simultaneously-presented stimuli, usually explained via divisive normalization \citep{yiltiz_contingent_2020,aschner_temporal_2018,solomon2014moving}, which is not included in our model (see Appendix~\ref{appendix:alt_models}).
One possible way to bridge this gap would be to combine our normative approach with recently-proposed recurrent models of normalization \citep{heeger_organics_2018,heeger_recurrent_2020}.

\paragraph{Alternative network architectures.}
There are alternative, equivalent formulations of our model that may give rise to the same steady-state responses as Eq.~\ref{eq:ss}, which we illustrate in Appendix~\ref{appendix:alt_models}.
Firstly, our model is equivalent to a two-stage feedforward network with gain modulation preceding the inputs of the second stage.
Since orientation tuning arises in V1, these two stages could be two different layers within V1; the core mechanism of our framework can thus be related to studies describing adaptive gain changes being inherited from one group of neurons to the next \citep{kohn_neuronal_2003, dhruv_cascaded_2014,stocker2009visual}.
Secondly, gain modulation in our model, which serves to multiplicatively scale input drive, $\rvf(\rvs)$, can equivalently be interpreted as multiplicatively \textit{attenuating} the recurrent drive of the network.
In this sense, our model resembles that of \citet{heeger_recurrent_2020}, in which divisive normalization is mediated by gating recurrent amplification.

\paragraph{Experimental predictions.}
We propose that rapid neural population adaptation in cortex can be mediated by single neuron adaptive gain modulation. 
Validating this hypothesis would require careful experimental measurements of neurons during adaptation. 
First, our framework predicts that between-neuron synaptic connectivity (i.e. $\rmW$) remains stable through adaptation.
Second, our normative objective suggests that gain homeostasis plays a central role in population adaptation (see Appendix~\ref{appendix:ablations}).
Evidence for stimulus-dependent gain control such as this can possibly be found by measuring neuron membrane conductance during adaptation, mediated by changes in slow hyperpolarizing Ca\textsuperscript{2+}- and Na\textsuperscript{+}-induced K\textsuperscript{+} currents \citep{sanchez2000membrane}.
Lastly, while there has been considerable progress in mapping the circuits involved in sensory adaptation \citep{wanner_whitening_2020}, determining the exact structure of functional recurrent connectivity remains an open problem.  
Indeed, we show how different (but not all) forms of $\rmW$ can give rise to the same qualitative results shown here (Appendix~\ref{appendix:W}).
Performing adaptation experiments with richer sets of stimulus ensembles, $p(\rvs)$, can provide better constraints for solving this functional inverse problem.

\subsection{Conclusion}

We demonstrate that adaptation effects observed in cortex -- changes in response maxima and minima, tuning curve repulsion, and stimulus-dependent response decorrelation -- can be explained as arising from the recurrent propagation of single neuron gain adjustments aimed at coding efficiency.  
This adaptation mechanism is general, and can be applied to modalities other than vision. 
For example, studies of neural adaptation in auditory cortex have shown that adaptive responses such tuning curve shifts cannot be explained by feedforward mechanisms, and likely arise from adaptive changes to intracortical recurrent interactions \citep{hershenhoren_intracellular_2014,lohse2020neural}.
Previous population adaptation models rely on changes in all-to-all synaptic weights to explain these phenomena \citep[e.g.][]{westrick_pattern_2016}, but our results suggest that single neuron gain modulations may provide a more plausible mechanism which uses $\mathcal{O}(N)$ instead of $\mathcal{O}(N^2)$ adaptive parameters. 
Adaptation in cortex happens on the order of hundreds of milliseconds, and is just as quickly \textit{reversible} \citep{muller1999rapid}; a network whose synaptic weights were constantly remapping would be undesirable due to a lack of stability, while a mechanism such as adaptive single neuron gain modulation can be local, fast, and reversible \citep{ferguson_mechanisms_2020}.
Taken together, our study offers a simple mechanistic explanation for observed adaptation effects at the level of a neural population, and expands upon well-established concepts of adaptive coding efficiency with single neuron gain control.

\clearpage





\begin{ack}

We thank Matteo Carandini for providing us with V1 neural recording data.
We also thank Teddy Yerxa, Pierre-Étienne Fiquet, Stefano Martiniani, Shivang Rawat, Gabrielle Gutierrez, Ann Hermundstad, and Wiktor Młynarski for their feedback on earlier versions of this work.

\end{ack}


{ \small
\bibliography{refs.bib}
}

\newpage
\appendix

\counterwithin{figure}{section}
\counterwithin{equation}{section}
\pagenumbering{arabic}  

\section{Details on model recurrent connectivity matrix}
\label{appendix:W}
\subsection{Initializing the recurrent connectivity matrix $\rmW$}
We restrict ${\rmW} \in \mathbb{R}^{N \times N}$ to the space of circularly symmetric (i.e. convolutional) positive definite matrices.
In our model, the recurrent weight kernel forming the convolutional matrix is (net) positive everywhere, with higher probability between similarly tuned excitatory neurons than between dissimilarly tuned neurons \citep{ko2011functional,lee2016anatomy}.
The kernel we use is a Gaussian (10\textsuperscript{$\circ$} FWHM) summed with a uniform density.
To prevent recurrence from diverging (Eq.~\ref{eq:ss}), the operator norm of $\rmW$ (i.e. the max eigenvalue) must be less than 1.
We fixed $\Vert \rmW \Vert_\text{Op}=0.8$ for all recurrent weight matrices in this study.

\subsection{Recurrent synaptic connectivity influences adaptation effects}\label{appendix:recurrent_contributions}

\begin{figure}[htb]
\centering
\includegraphics[width=.6\textwidth]{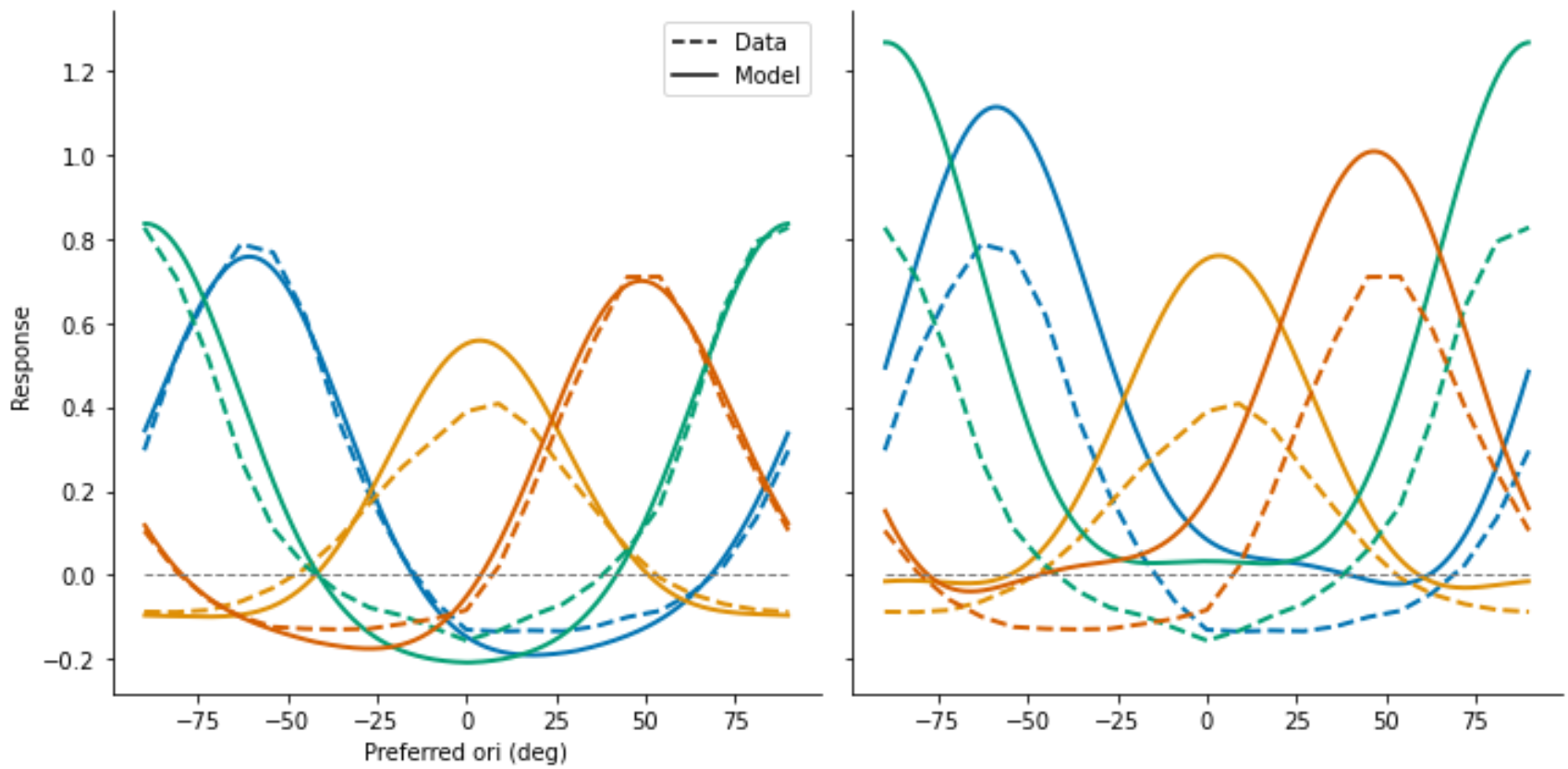}
\caption{
\small
Data (dashed) vs model (solid) with different forms of $\rmW$.
Dashed lines (identical in left and right panels) are observed post-adaptation response curves for a subset of the population.
{\bf Left} Model from main text, using a convolutional $\rmW$ with recurrent net excitation from similarly tuned  neurons, and broad/untuned net excitation from dissimilarly tuned neurons.
{\bf Right} Simulated post-adaptation responses from a model with $\rmW$ comprising recurrent net excitation from similarly tuned neurons, and net \textit{inhibition} from dissimilarly tuned neurons.
}
\label{fig:w_effects}
\end{figure}

The structure of the recurrent weight matrix $\rmW$ greatly impacts the adaptive changes in neural responses.
Specifically, we find that, with our recurrent model and objective (Eq.~\ref{eq:ss} and Eq.~\ref{eq:objective}), weak net excitatory inputs from dissimilarly tuned neurons is needed to capture the observed effects in data.
This recurrent composition is line with broad, untuned excitatatory signal amplification contributing to overall background activity in cortical circuits \citep{reinhold_distinct_2015}.
For reference, we reproduce a subset of the post-adaptation tuning curves from the main text here in Fig.~\ref{fig:w_effects}A.

\begin{figure}[htb]
\centering
\includegraphics[width=\textwidth]{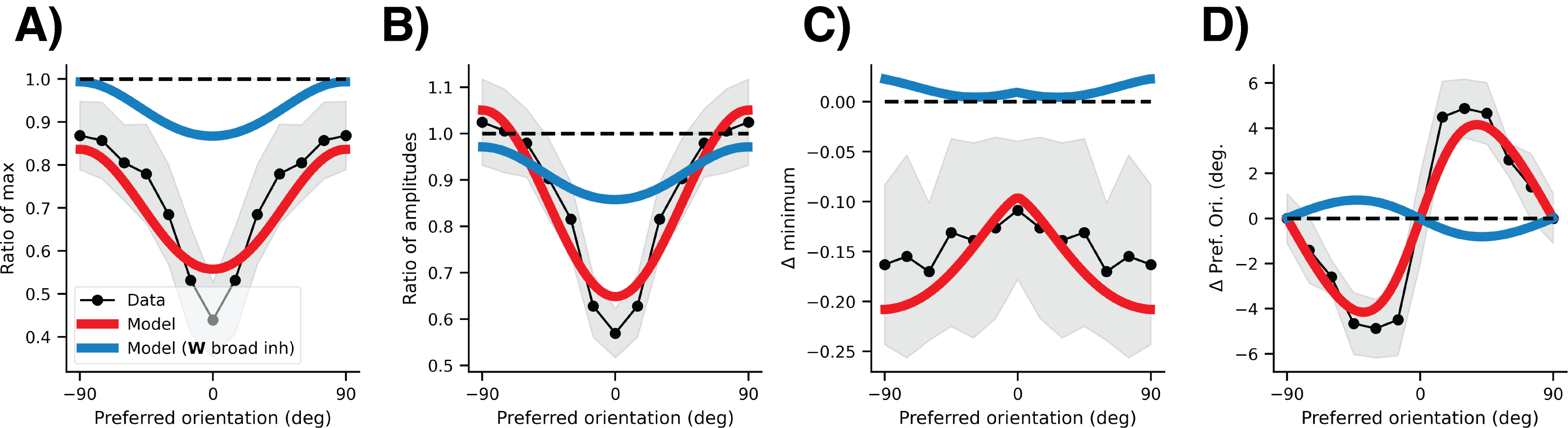}
\caption{\small
All panels are the same as Fig.~\ref{fig:tc_changes} in the main text, but blue is now a model with $\rmW$ set to a convolutional matrix with a Mexican hat kernel.
Notably, this model cannot reproduce the adaptive maxima, minima, and amplitude effects observed in data.
Furthermore, the tuning curves are no longer repelled from the adapter (panel {\bf D)}, but are instead \textit{attracted} toward it.
}
\label{fig:tc_broad_inh_stats}
\end{figure}

Figure~\ref{fig:w_effects}B shows response curves from an example model using a convolutional $\rmW$ derived from the Mexican hat weight kernel, which is an excitatory Gaussian (10\textsuperscript{$\circ$} FWHM) minus a wider Gaussian (60\textsuperscript{$\circ$} FWHM) \citep[][]{carandini_predictions_1997,teich2010v1,quiroga_adaptation_2016}.
We re-scaled the weight matrix to have operator norm of 0.8.
The responses in Fig.~\ref{fig:w_effects}B here are from a model minimizing $\ell_2$ error between the data and model adapted responses after a hyper-parameter sweep ($\alpha=$ 3E-4, $\gamma=$ 2E-3).
In contrast to the recurrent weight matrix we use in the main text (described above), the Mexican hat kernel has recurrent net excitation from neurons with similar tuning, and \textit{net inhibition} from neurons with dissimilar tuning.
A model with this recurrent weight kernel is unable to capture the effects of response maxima, minima, and amplitudes, and produces tuning curve \textit{attraction} rather than repulsion from the adapter Fig.~\ref{fig:tc_broad_inh_stats}A-D.
Taken together, this suggests that broad, untuned weak recurrent excitation is necessary for our model to capture the wide array of post-adaptation effects found in this dataset. 

\section{Objective ablation}
\label{appendix:ablations}
Here, we assess the contribution of each term of the objective  (Equation~\ref{eq:objective}) to explain the adaptation effects found in data.

\subsection{Gain homeostasis confers representation stability with adaptation}
\begin{figure}[htb]
    \centering
    \includegraphics[width=.6\textwidth]{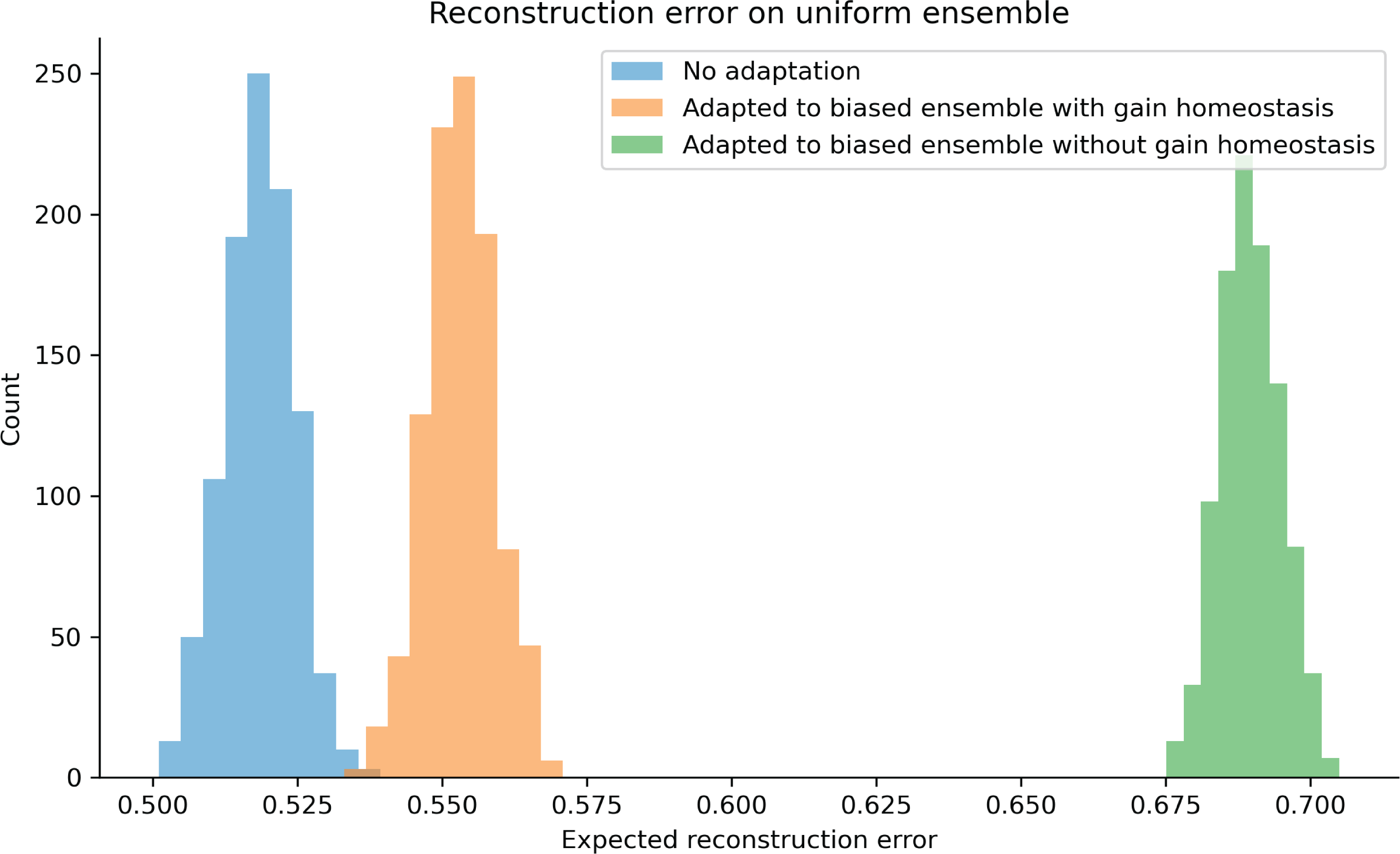}
    \caption{\small Gain homeostasis induces stability across statistical contexts. 
    Histograms are bootstrap samples (1000 repeats) of the average stimulus reconstruction error under the uniform stimulus ensemble without adaptation, after adaptation with gain homeostasis, and after adaptation without gain homeostasis. 
    }
    \label{fig:gain_homeo_reconstruction}
\end{figure}

Fig.~\ref{fig:gain_homeo_reconstruction} shows the impact of removing the gain homeostasis term from Eq~\ref{eq:objective}.
Gain homeostasis prevents the network from drastically re-configuring its representation after adaptation, and allows the network to maintain a stable representation of the stimulus ensemble (compare orange to green).

\subsection{Contributions of each term to adaptation}
\begin{figure}[htb]
    \centering
    \includegraphics[width=\textwidth]{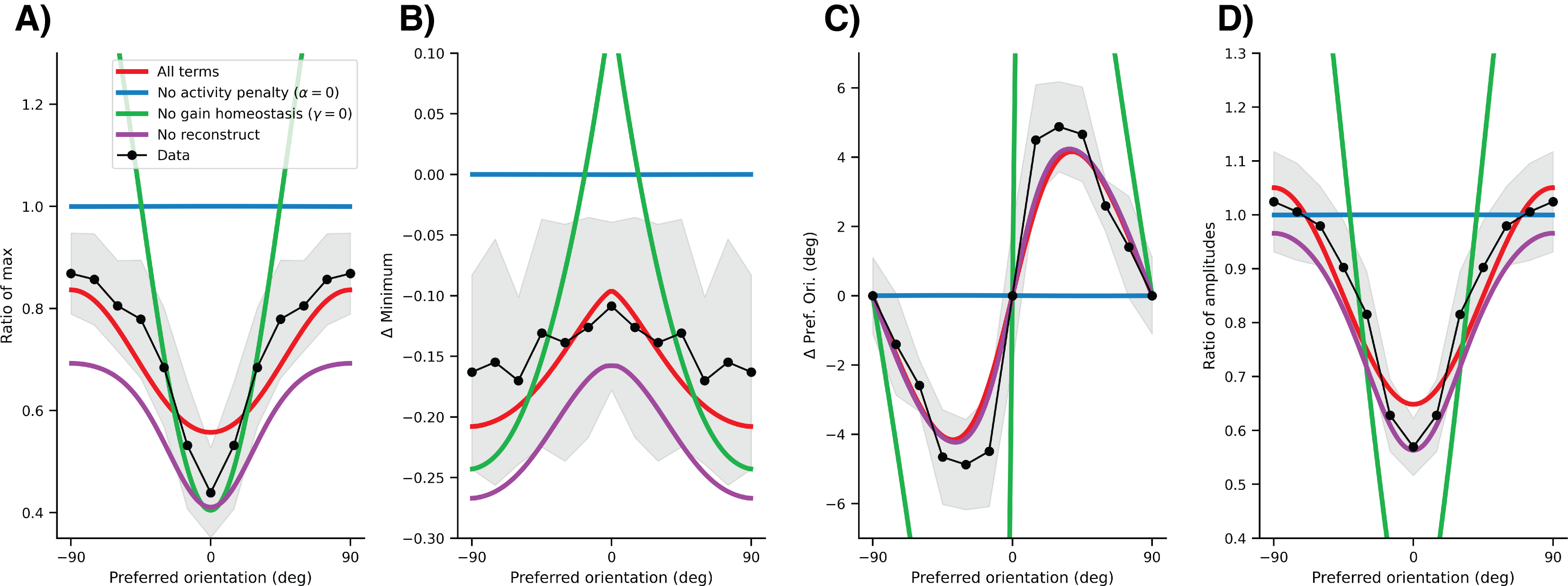}
    \caption{\small Each term in the objective (Equation~\ref{eq:objective}) is necessary to account for the full array of adaptation effects observed in data.}
    \label{fig:objective_ablation}
\end{figure}

We assess the importance of the three terms in the objective (Eq.~\ref{eq:objective}) and show that they are all jointly necessary to produce the effects shown in main text.
Figure~\ref{fig:objective_ablation} shows the adapted model responses using Eq.~\ref{eq:objective} to adapt in red.
Without the gain homeostasis term (i.e. $\gamma=0$, green), the gains radically change after adapting to the biased stimulus ensemble.
This produces higher responses in neurons tuned for orientations along the flank (far from the adapter at zero degrees), and completely fails to reproduce any of the adaptation effects observed in data.
Without the activity penalty (i.e. $\alpha=0$, blue), the model's responses are equivalent to one with no adaptation.
Finally, without the $\ell_2$ reconstruction penalty (first term of objective; purple), the maxima and minima undershoot what is observed in data; however, the model does reasonably well at capturing the shifts in tuning preference and response amplitude.
The reconstruction term of the objective encourages the network to maintain a high fidelity representation of the stimulus after adaptation.
This ablation finding suggests that the shifts in tuning preference observed in many previous studies \citep{clifford_visual_2007} may arise from adaptive sensory information-preserving properties of the system.
Taken together, each component of the objective works in concert to yield the adaptation response phenomena seen in th data.

\section{Analytic solution to the adaptation objective}
\label{appendix:isolating_g}
From Eq. \ref{eq:ss}, the steady state response of a network is given by

\begin{align} 
\rvr_*({\bf s}, {\bf g}) &= [{\bf I} - {\rmW}]^{-1} \left({\bf g} \circ {\bf f}({\bf s})\right) \nonumber \\
\rvr_*({\bf s}, {\bf g}) &= {\bf M} \left({\bf g}\circ{\bf f}({\bf s})\right) \label{eq:ss_simplified}
\end{align}
where ${\rmW} \prec {\bf I}$, and ${\bf M} := [{\bf I} - {\rmW}]^{-1}$ is a matrix capturing the effect of leak and lateral recurrence in the network. 
The feedforward and recurrent weights of our model are assumed to be fixed through adaptation.
We can isolate ${\bf g}$ using the identity $\diag{\bf a}{\bf b} = \diag{\bf b}{\bf a}$, for two vectors $\bf a$ and $\bf b$, to get:

\begin{align*}
 {\bf M} \left( {{\bf g}} \circ {\bf f}({\bf s}) \right) &= {\bf M} \diag{{\bf f}({\bf s})} {\bf g}\\
  &\equiv {\bf H}({\bf s}){\bf g},
\end{align*}
where we define ${\bf H}: \mathbb{R}^N \mapsto \mathbb{R}^{N\times N}$ as a linear operator that maps {\bf s} to a matrix using {\bf M} and ${\bf f}({\bf s})$.

In the main text, the loss functional (Eq.~\ref{eq:objective}) omitted dependence on the decoder ${\bf D}$ for clarity.
The full objective is

\begin{align}
\mathcal{L}(p({\bf s}), {\bf g}, {\bf D})&= \mathbb{E}_{{\bf s} \sim p({\bf s})} \left[\underbrace{\parallel {\bf s} - {\bf D}^\top {\bf H}({\bf s}){\bf g}\parallel_2^2}_{\text{reconstruction}} + \alpha \underbrace{\parallel {\bf H}({\bf s}){\bf g} \parallel_2^2}_{\text{activation}} \right] + \gamma \underbrace{ \parallel {\bf g} - {\bf g}_0\parallel^2_2 }_{\text{homeostatic gain}} + \delta \underbrace{\parallel {\bf D} \parallel_{\text{F}}^2}_{\text{decoder}},
\end{align}
where $\delta$ is a hyperparameter controlling the decoder weights $\rmD$.
The results in the main text have $\delta$ set to zero, and our findings do not qualitatively change with small deviations away fro $\delta=0$.
This objective is bi-convex in ${\bf D}$ and ${\bf g}$ (i.e. convex when one of the two variables is held fixed).
Indeed, with ${\bf g}$ fixed, the loss simply becomes an $\ell_2$-regularized least-squares problem in ${\bf D}$.
One can show that the linear decoder regularization term, $\delta$, is equivalent to assuming noisy outputs with additive isotropic Gaussian noise.
We solve for each optimization variable in alternation until they reach convergence.
As our stimulus ensemble comprises a discrete set of $K$ stimuli, we can write our objective explicitly as a weighted summation,
\begin{equation}
\mathcal{L}({\bf g})= \frac{1}{2} \sum_{k=1}^K p({\bf s}_k) \left\{  \parallel {\bf s}_k - {\bf D}^\top {\bf H}({\bf s}_k){\bf g}_s\parallel_2^2 + \alpha \parallel {\bf H}({\bf s}_k){\bf g} \parallel_2^2\right\} + \gamma \parallel {\bf g} - {\bf g}_0\parallel^2_2 .
\end{equation}

Computing $\nabla \mathcal{L}_{\bf g}=0$, and isolating for ${\bf g}$ yields a linear system of equations,
\begin{equation}
\left [\sum_k p({\bf s}_k) \left\{ {\bf H}({\bf s}_k)^\top \left({\bf D D}^\top + \alpha {\bf I}\right){\bf H}({\bf s}_k ) \right\} + \gamma {\bf I} \right] {\bf g} = \left[ \sum_k p({\bf s}_k)\left\{ {\bf H}({\bf s}_k)^\top {\bf D} \hat{\bf s}_k\right\} \right] + \gamma {\bf g}_0.
\end{equation}
This is in the form  of $ {\bf A g} = {\bf b}$ and can therefore be solved exactly (e.g. using \texttt{numpy.linalg.solve()}).
A similar derivation can be done for the optimal ${\bf D}$.
To initialize ${\bf g}_0$ and ${\bf D}$, we alternated optimization between ${\bf D}$ and ${\bf g}$ (using the control context stimulus ensemble) until convergence using co-ordinate descent.

\section{Alternative models}
\label{appendix:alt_models}

\subsection{Equivalent circuit with adaptive recurrent gain}
Here, we explore an alternative network parameterization which has identical steady-state behavior as the network we have selected for our model: consequently, adaptation under our training procedure will have identical behavior at a network level for both parameterizations. The dynamics of our network, replicated from Equation \ref{eq:drdt_linear} for clarity, are given by

\begin{align}
     \frac{d\rvr(\rvs, \rvg)}{dt} &= -\rvr + {\rmW}  \rvr + {\bf g} \circ {\bf f}({\bf s}) , \nonumber  \\
    &= [-{\bf I + W}]\rvr + {\bf g} \circ {\bf f}({\bf s}), 
\end{align}
where we denote both the leak term,$-{\bf I r}$, and the recurrent weight term, ${\rmW} \rvr$, as the recurrent drive.
As an alternative to this parameterization, consider instead multiplicatively scaling each neuron's recurrent drive with ${\bf g} \in \mathbb{R}^N_+$,
\begin{equation}
     \frac{d\rvr(\rvs, \rvg)}{dt} = {\bf g}^{-1}\circ[-{\bf I+W}]\rvr + {\bf f}({\bf s}), \label{eq:drdt_g2}
\end{equation}
where ${\bf g}^{-1} = [1/g_1,1/g_2,...,1/g_N]^\top$.
Intuitively, Equation~\ref{eq:drdt_g2} states that an increase in each neuron's $g_i$ attenuates its recurrent drive.
Gain changes in this model adjust the overall sensitivity to recurrent (including self-recurrence, i.e. leak) drive.
To solve for this new network's steady-state, $\rvr_\ast({\bf s}, {\bf g})$, we set Equation \ref{eq:drdt_g2} to zero and solve,

\begin{align}
      {{\bf g}^{-1}}\circ[{\bf I-W}] \left(\rvr_\ast({\bf s}, {\bf g})\right) &=  {\bf f}({\bf s}) \\
     \rvr_\ast({\bf s}, {\bf g}) &= \left[ \diag{{\bf g}^{-1}}[{\bf I-W}] \right]^{-1} {\bf f}({\bf s}) \\
     \rvr_\ast({\bf s}, {\bf g}) 
     &= [{\bf I-W}]^{-1} \left({{\bf g}} \circ {\bf f}({\bf s})\right), \nonumber
\end{align}
which is the same as our steady-state equation given by Equation \ref{eq:ss} in the main text.
Therefore, our original formulation of multiplicatively scaling the network's feedforward drive is mathematically equivalent to inversely scaling (attenuating) its recurrent drive.
This means that upscaling gain on feedforward inputs is equivalent to downscaling net inhibition on each neuron.

\subsection{Equivalent circuit with two-layer feedforward architecture}

The overall action of the network at steady-state (Equation \ref{eq:ss}) is to mix the gain-modulated feedforward responses, ${\bf g \circ f(s)}  $, by a linear transformation dependent on the recurrent circuitry $[\rmI - \rmW]^{-1}$.
The steady-state response of this recurrent network is \textit{equivalent} to a two-layer feedforward network with gain modulation after the first layer.
This means that viewing the steady-state responses alone, as is frequently done in neurophysiological adaptation experiments \citep{weber_coding_2019}, it is impossible to tell whether the system was exclusively feedforward or recurrent.
This is well-known \citep{dayan2005theoretical} and is a fundamental result in signal processing theory of linear feedback systems.
Our model can be interpreted as a cascade of transformations, propagating adaptive response changes downstream \citep{kohn_neuronal_2003, dhruv_cascaded_2014}. 

\subsection{Relationship to divisive normalization}

Our model is linear and \textit{not} divisive.
Writing the steady-state for the $i$\textsuperscript{th} neuron explicitly (omitting ${\bf g}$ to reduce clutter) yields

\begin{align*}
   r_i &=  f_i({\bf s}) + \sum_{j=1}^N w_j r_j \\
   r_i &= f_i({\bf s}) + \sum_{j\neq i} w_j r_j + w_i r_i\\
   (1-w_i)r_i &=  f_i({\bf s}) + \sum_{j\neq i} w_j r_j \\
   r_i &= \frac{1}{1 - w_i} \left[ \sum_j f_j ({\bf s}) + \sum_{j\neq i} w_j r_j \right].
\end{align*}
Thus, there is no nonlinear interaction between $r_i$ and other neurons in the population.

\end{document}